\documentclass[12pt]{iopart}

\usepackage{graphicx}
\begin{document}
\title{Degeneracy of Majorana bound states and fractional Josephson effect in a dirty SNS junction}

\author{S. Ikegaya$^{1}$, and  Y. Asano$^{1,2,3}$}

\address{$^{1}$ Department of Applied Physics,Hokkaido University, Kita 13 Nishi 8, Sapporo 060-8628, Japan}
\address{$^{2}$ Center of Topological Science and Technology,Hokkaido University, Kita 13 Nishi 8, Sapporo 060-8628, Japan}
\address{$^{3}$ Moscow Institute of Physics and Technology, 141700 Dolgoprudny, Russia}
\ead{satoshi-ikegaya@eng.hokudai.ac.jp}
\begin{abstract}
We theoretically study the stability of more than one Majorana Fermion appearing in a
$p$-wave superconductor/dirty normal metal/$p$-wave superconductor junction in two-dimension
by using chiral symmetry of Hamiltonian.
At the phase difference across the junction $\varphi$ being $\pi$, we will show that
all of the Majorana bound states in the normal metal belong to the same chirality.
Due to this pure chiral feature, the Majorana bound states retain their high degree 
of degeneracy at the zero energy even in the presence of random potential.
As a consequence, the resonant transmission of a Cooper pair via the degenerate MBSs
carries the Josephson current at $\varphi=\pi-0^+$, which explains
the fractional current-phase relationship discussed in a number of previous papers.
\end{abstract}
%
%

\section{Introduction}

Exotic properties of Majorana fermions (MFs)\cite{mj37} is a hot issue in condensed matter physics.
MFs emerge as the surface bound states of 
 topologically nontrivial superconductors such as 
$p$-wave superconductors \cite{rg00,yk01}, topological insulator/superconductor heterostructures \cite{fk08},
spin-orbit coupled semiconductor/superconductor heterostructures \cite{ms09,js10,ja10,rl10,yo10,jy13}
and Shiba chains \cite{tc11,np13}.
Such the Majorana fermion bound states (MBSs) have attracted much attention 
from a view of the fault-tolerant topological quantum computation \cite{di01,js11}.
Thus, the realization of MBSs is a recent desired subject in experimental fields \cite{mourik,deng,yazdani}.
Since a Majorana fermion corresponds to a half of an electron, MFs 
 always emerge as pairs spatially separated from each other.
When a MF stays at one edge of a superconductor, its partner stays at the other edge.
Generally speaking, more than one MF staying at the same place are unstable 
because they may couple back to an electron.

When two one-dimensional semi-infinite $p$-wave superconductors  
are joined in a superconductor/insulator/superconductor (SIS) junction,
a pair of MFs staying at the two junction interfaces 
form the Andreev bound states. As a consequence, 
the Josephson current exhibits 
the fractional current-phase ($J$-$\varphi$) relationship of $J(\varphi) \propto \sin (\varphi/2)$ at the zero temperature \cite{yk01,kwon}.
The fractional Josephson effect is especially important because the effect 
provides a direct way of read-out process in the fault-tolerant topological computation \cite{js11}.
Here we note that $J$ is always $2\pi$ periodic in the direct-current Josephson effect.
Thus the fractional current-phase relationship (CPR) means that 
the current jumps at $\varphi=\pm \pi$.
It has been well known that ballistic junctions~\cite{ishii,ishii2,ko2,likharev} and 
SIS junctions with unconventional pairing symmetries~\cite{buchholts,tanaka0,hu,ya04-1,yt96-1,barash} also indicate the fractional CPR.
The unique feature to $p$-wave junctions is the persistence of the fractional CPR even
in the presence of random impurity potential~\cite{ya13}.
In fact, a theoretical study~\cite{ya06,ya062} reported the fractional Josephson effect
in a two-dimensional $p_x$-wave superconductor/dirty normal metal/$p_x$-wave superconductor (SNS) junction. 
More than one MF are degenerate at the zero-energy in the dirty normal metal
and induce the resonant transmission of the Cooper pair at $\varphi=\pi-0^+$.
Generally speaking, the large degree of degeneracy in quantum states is a result of high symmetry of Hamiltonian.
However it has been unclear what symmetry protects the degeneracy of the MBSs in a dirty normal metal. 
We address this issue in the present paper.

Several previous studies have suggested that chiral symmetry of Hamiltonian is a key feature 
to explain the stability of more than one MF at a surface of topologically nontrivial 
superconductors~\cite{tewari,niu,diez,si15}.
On the basis of these novel insight,
we will prove the robustness of the degenerate MBSs in diffusive SNS junctions. In addition, we reconsider 
the meaning of a phenomenological theory of the fractional Josephson effect, where
the tunneling Hamiltonian between the two edges at either sides of the insulator is described by 
$H_T= -i t \cos(\varphi/2) \gamma_L \gamma_R$~\cite{yk01,kwon}. 
Here $t$ is the tunneling amplitude and $\gamma_L$ ($\gamma_R$) is the operator of a MF at the edge 
of the superconductor on the left (right)-hand side of the insulator.
The Josephson current calculated from $J \propto \partial_\varphi \left\langle H_T \right\rangle$ exhibits the fractional 
CPR. However, this argument may be self-contradicted.
The Josephson current flows at $\varphi=\pi-0^+$ while the tunneling Hamiltonian vanishes.
We also try to solve this puzzle in the present paper.  

\section{chiral symmetry} 
\begin{figure}[hhhh]
\begin{center}
\includegraphics[width=0.5\textwidth]{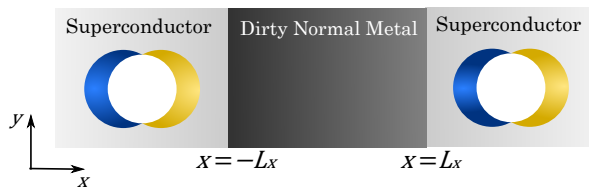}
\caption{Schematic image of
the $p_x$-wave superconductor/dirty normal metal/the $p_x$-wave superconductor junction.}
\label{fig:junct}
\end{center}
\end{figure}
Let us consider a two-dimensional SNS junction where two superconductors are characterized by 
an equal-spin-triplet $p_x$-wave symmetry as shown Fig.~\ref{fig:junct}.
The junction consists the three segments:
a dirty normal metal ($-L_x \leq x \leq L_x$),
and two superconductors ($L_x \leq j \leq \infty$
and $-\infty \leq j \leq - L_x$).
The junction is described by the Bogoliubov-de Gennes Hamiltonian
\begin{eqnarray}
H (\varphi ) = H_L+ H_N + H_R(\varphi),\\
H_L= \left[
\begin{array}{cc}
\xi (\bi{r})
&  \frac{ \Delta }{k_F} \partial_x \\
-\frac{ \Delta }{k_F} \partial_x
& - \xi (\bi{r}) \\
\end{array}
\right], \label{eq:hml}\\
H_R(\varphi)=
\left[
\begin{array}{cc}
\xi (\bi{r})
&  \frac{ \Delta e^{i\varphi} }{k_\mathrm{F}} \partial_x \\
-\frac{ \Delta e^{-i\varphi} }{k_\mathrm{F}} \partial_x
& - \xi (\bi{r}) \\
\end{array}
\right], \label{eq:hmr}\\
H_N= \left[
\begin{array}{cc}
\xi (\bi{r}) + V_{\mathrm{imp}}(\bi{r})
& 0 \\
0
& - \xi (\bi{r}) - V_{\mathrm{imp}}(\bi{r})\\
\end{array}
\right],\\
\xi (\bi{r}) = 
- \frac{\hbar^2}{2m} \nabla^2 - \mu, \qquad
k_{\mathrm{F}} = \sqrt{2m \mu} / \hbar,
\end{eqnarray} 
where $m$ denotes the effective mass of an electron,
$\mu$ is the chemical potential,
and $\Delta$ denotes the amplitude of the pair potential.
In what follows, we consider $2\times 2$ BdG Hamiltonian for one spin sector.
The phase difference between the two superconductors is denoted by $\varphi$.
The random impurity potential in the normal segment is 
represented by $V_{\mathrm{imp}} (\bi{r})$.

It is easy to confirm the following relations,
\begin{eqnarray}
\Gamma H_L \Gamma^{-1}= - H_L, \label{cl}\\
(\rme^{\rmi\varphi T_3} \Gamma)\;  H_R \; (\rme^{\rmi\varphi T_3} \Gamma)^{-1}= - H_R, \label{cr}\\
\Gamma H_N \Gamma^{-1}= - H_N, \label{cln} \\
 (\rme^{\rmi\varphi T_3} \Gamma) \; H_N \; (\rme^{\rmi\varphi T_3} \Gamma)^{-1}= - H_N, \label{crn}\\
\Gamma =\left[
\begin{array}{cc}
0 & 1 \\
1 & 0 \\
\end{array} \right] = T_1, \qquad 
T_3 =\left[
\begin{array}{cc}
1 & 0 \\
0 & -1 \\
\end{array} \right],
\end{eqnarray}
Eq.~(\ref{cl}) represents chiral symmetry of $H_L$ with respect to $\Gamma$.
In the same way, Eq.~(\ref{cr}) represents chiral symmetry of $H_R$ with respect to $\rme^{\rmi\varphi T_3} \Gamma$.
The Hamiltonian in the normal part $H_N$ preserves chiral symmetry for both $\Gamma$ and $\rme^{\rmi\varphi T_3} \Gamma$. 
When a Hamiltonian preserves chiral symmetry, the eigenstates of the Hamiltonian have two important features~\cite{ms11}.
In the case of Eq.~(\ref{cl}), for instance, one can prove following properties of eigen states of $H_L$.

\noindent (i) The eigenstates of the $H_L$ at the zero energy 
are simultaneously the eigenstates of $\Gamma$ with its eigenvalue (chirality) either $\gamma=+1$ or $-1$.

\noindent (ii) On the other hand, the nonzero-energy states of $H_L$ are described by the linear combination of
two different eigenstates of $\Gamma$: 
one has $\gamma=+1$ and the other has $\gamma=-1$.

\noindent Below we prove the stability of the highly degenerate zero energy states
appearing in the SNS junction by taking these features into account.
We note that the total Hamiltonian $H$ preserves $\Gamma\; H \Gamma^{-1}=-H$ 
for $\varphi$ being either 0 or $\pm \pi$.

We first analyze the chiral property of the zero-energy states appealing at the surface of the two
semi-infinite superconductors ($x \leq -L_x$ and $x \geq L_x$).
To do this, we remove the normal segment ($-L_x \leq x \leq L_x$) and apply the hard-wall boundary condition 
at $x=-L_x$ and $x=L_x$.
In the $y$ direction,
the width of the superconductors is $W$ and the hard-wall boundary condition is applied.
By solving the Bogoliubov-de Gennes equation,
we obtain the wave function for the the zero-energy states as
\begin{eqnarray}
\psi_{L,n}(\bi{r}) 
= \frac{1}{\sqrt{N_n}}
\left[
\begin{array}{cc} 1 \\ 1 \\
\end{array}
\right]
X_{n,+} (x) Y_n(y),
\label{eq:zes-l} \\
\psi_{R,n}(\bi{r}) 
= \frac{1}{\sqrt{N_n}}
\left[
\begin{array}{cc} \rme^{\rmi \frac{\varphi}{2}} \\ -\rme^{-\rmi \frac{\varphi}{2}}
\end{array}
\right] X_{n,-}(x) Y_n(y), \\
\label{eq:zes-r}
X_{n,\pm}(x) = {\rm sin}[ q_n (x \pm L_x)] \rme^{\pm x/\xi}, \\
Y_n(y) = \sqrt{\frac{2}{W}} \sin \left( \frac{n \pi}{W} y \right),\label{yn}\\
q_n = \sqrt{ k_n^2-\xi^{-2}}, \quad \xi=\hbar^2 k_{\mathrm{F}}/m \Delta_0, \\
k_n = \frac{\sqrt{2 m \mu_n} } {\hbar}, \qquad
\mu_n= \mu-\frac{\hbar^2}{2m} \left( \frac{n \pi}{W} \right)^2,
\end{eqnarray}
where $n$ indicates the propagating channels.
The wave function $\psi_{L,n} $ ($\psi_{R,n} $) represents
the $n$-th zero-energy state localized at the surface of
the left (right) superconductor.
The normalization coefficient is denoted by $N_n$.
The degree of degeneracy at the zero energy is equal to the number of the propagating channels $N_c$ 
because a zero-energy state can be defined for each propagating channel.
The derivations of the wave functions are shown in \ref{sec:zs}.
As indicated by the property (i), the zero-energy states in Eqs.~(\ref{eq:zes-l}) and (\ref{eq:zes-r}) are the 
eigenstates of $\Gamma$ and $\rme^{\rmi\varphi T_3} \Gamma$, respectively.

The particle-hole symmetry of the total Hamiltonian is represented by
\begin{eqnarray}
\Xi \; H \; \Xi^{-1} = - H, \\
\Xi = \Gamma \; \cal{K},
\end{eqnarray}
where $\cal{K}$ denotes the complex conjugation. 
Since
\begin{eqnarray}
\Xi \psi_{L,n} = \psi_{L,n},\\
\Xi \psi_{R,n} = - \psi_{R,n},
\end{eqnarray}
all the zero-energy states are the Majorana bound states.
Thus, at a surface of a $p_x$-wave superconductor, the degree of the degeneracy in MBSs is $N_c$.

\section{zero-energy states in SNS junctions}
\begin{figure}[tttt]
\begin{center}
\includegraphics[width=0.5\textwidth]{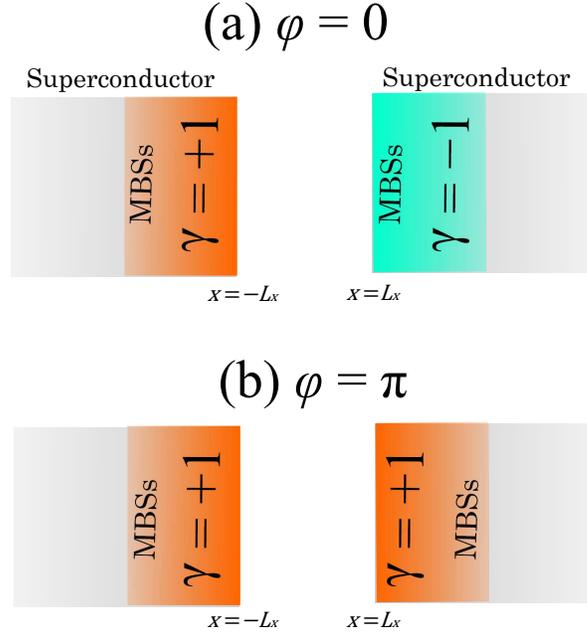}
\caption{Schematic image for the chirality of the Majorana bound states. 
(a) At $\varphi=0$, the left-side MBSs and the right-side MBSs have the opposite chirality each other.
(b) On the other hand, at $\varphi= \pi$, both left-side and right-side MBSs have the chirality $\gamma=+1$.}
\label{fig:mf}
\end{center}
\end{figure}

To analyze the MBSs in a SNS junction, we insert a normal segment described by $H_N$
into the two superconductors.
At $\varphi=0$,
the wave function $\psi_{L,n}$ and $\psi_{R,n}$ satisfies
\begin{eqnarray}
\Gamma \psi_{L,n} &= \psi_{L,n}\\
\Gamma \psi_{R,n} &= - \psi_{R,n},
\end{eqnarray}
for all $n$. Namely, all the MBSs in the left superconductor belong to $\gamma=+1$
while those in the right superconductor belong to $\gamma=-1$
 as shown in Fig.~\ref{fig:mf} (a).
The MBSs at the surface of the two different superconductors 
have the opposite chirality to each other.
In a SNS junction, a normal metal connects the two superconductor. 
MBSs with $\gamma=+1$ (MBSs with $\gamma=-1$) penetrate into the normal metal from the 
left (right) superconductor. 
As a result, they form nonzero-energy states there. 
In this way, the penetration of MBSs into the normal metal lifts 
the high degeneracy at the zero-energy. 
In other wards, pairs of MFs couple back to electrons and the number of 
such pairs is $N_c$.

On the other hand
at $\varphi=\pi$, one can find
\begin{eqnarray}
\Gamma \psi_{L,n} = \psi_{L,n}\\
\Gamma \psi_{R,n} =  \psi_{R,n}.
\end{eqnarray}
Both $\psi_{L,n}$ and $\psi_{R,n}$ belong to the same chirality $\gamma=+1$ as shown in Fig.~\ref{fig:mf} (b).
The MBSs retain their high degree of degeneracy even in a SNS junction because 
the zero-energy states with $\gamma=-1$ are absent in the normal metal.
According to the property (ii), the zero-energy states belonging the same chirality cannot 
form any nonzero-energy states.

To confirm the argument above, we calculate the wave function in a SNS junction. 
We first set the impurity potential $V_{\mathrm{imp}} (\bi{r})=0$ 
and solve the Bogoliubov-de Gennes equation at the zero energy for $\varphi=\pi$, 
\begin{eqnarray}
H (\pi ) \psi_0 = 0.
\label{eq:bdgeq}
\end{eqnarray}
A solution of Eq.~(\ref{eq:bdgeq}) is given by (See also \ref{sec:zs})
\begin{eqnarray}
\psi_{L,n}^{\prime} (\bi{r}) =
\left[
\begin{array}{cc}
1 \\ 1 \\
\end{array}
\right]
\left[ a_{L} \rme^{\rmi q_n x} + b_{L} e^{- \rmi q_n x} \right] \rme^{x/\xi} Y_n (y),
\label{eq:wf-sl} \\
\psi_{N,n} (\bi{r}) =
\left[ \left[
\begin{array}{cc}
a_{N} \\ c_{N} \\
\end{array}
\right] \rme^{\rmi k_n x}
+
\left[
\begin{array}{cc}
b_{N} \\ d_{N} \\
\end{array}
\right] \rme^{- \rmi k_n x} \right] Y_n (y),
\label{eq:wf-n} \\
\psi_{R,n}^{\prime} (\bi{r}) =
\left[
\begin{array}{cc}
1 \\ 1 \\
\end{array}
\right]
\left[ a_{R} \rme^{\rmi q_n x} + b_{R} e^{- \rmi q_n x} \right] \rme^{-x/\xi} Y_n (y),
\label{eq:wf-sr}
\end{eqnarray}
where $\psi_{L,n}^{\prime}$,~$\psi_{N,n}$, and $\psi_{R,n}^{\prime}$ are the wave function 
at the $n$-th propagating channel in
the left superconductor, the normal metal, and the right superconductor, respectively.
By reflecting the chiral property of the MBSs in two superconductors,
the vector structure of the wave functions in the superconducting segments takes the particular form of
$\psi_{L(R),n}^{\prime} \propto [1,1]^{\rm T}$.
By applying the boundary condition at the two interfaces,
we obtain the two orthogonal zero-energy states for each propagating channel as
\begin{eqnarray}
\psi_{\pm} = \frac{1}{\sqrt{N_{\pm}}}
\left[
\begin{array}{cc}
1 \\ 1 \\
\end{array}
\right]
\phi_{n,\pm} (x) Y_n (y), \label{eq:wf-zes}  \\
\phi_{n,+} (x) = \cases{
A_+ (x) &for $x \leq -L_x$ \\
%
\sin (k_n x) &for $-L_x \leq x \leq L_x$ \\
%
%
B_+ (x) &for $x \geq L_x,$ \\} \\
%
%
\phi_{n,-} (x) = \cases{
A_- (x) &for $x \leq -L_x$ \\
%
\cos (k_n x) &for $-L_x \leq x \leq L_x$ \\
%
%
B_- (x) &for $x \geq L_x,$ \\} \\
A_{\pm} (x) = c_{\pm} \sin \{q_n (x+L_x) \mp \theta_{\pm} \} \rme^{(x+L_x)/\xi}, \\
B_{\pm} (x) = c_{\pm} \sin \{ q_n (x-L_x) \pm \theta_{\pm} \} \rme^{-(x-L_x)/\xi}, \\
c_{\pm} = \frac{\sqrt{k_n \{ k_n \pm \xi^{-1} \sin (2 k_n L_x)\}}} {q_n}, \\
\theta_+ = \arctan \left[ 
\frac{q_n \sin(k_n L_x)} {k_n \cos (k_n L_x) + \xi^{-1} \sin (k_n L_x)} \right] , \\
\theta_- = \arctan \left[
\frac{q_n \cos(k_n L_x)} {k_n \sin (k_n L_x) - \xi^{-1} \cos (k_n L_x)} \right],
\end{eqnarray}
where $N_{\pm}$ is a normalization coefficient.
Since we obtain the two zero-energy states for each propagating channel,
the degeneracy of the zero-energy bound states becomes twice the number of the propagating channel $N_c$.
More importantly, Eq.~(\ref{eq:wf-zes}) suggests that all the zero-energy states in SNS junction 
are the eigenstates of $\Gamma$ belonging to $\gamma=+1$.

Next we introduce the impurity potential $V_{\mathrm{imp}}$ into the normal segment. 
The random potential modifies the wave function in Eq.~($\ref{eq:wf-zes}$). 
Actually we cannot analytically describe how the wave function depends on
$\bi {r}$ anymore. But the 
vector part of the wave function $[1,1]^{\rm T}$ remains unchanged even in the presence of 
impurity potentials because $V_{\mathrm{imp}}$ preserves the chiral symmetry.
Therefore all the zero-energy states keep their chirality at $\gamma=+1$ 
even in the presence of $V_{\mathrm{imp}}$.
According to the property (ii), 
such chirality aligned zero-energy states keep their high degeneracy  
because they cannot construct nonzero-energy states in the absence 
of their chiral partner belonging to $\gamma=-1$.
As a result, the degenerate MBSs form the resonant transmission channels in the normal metal. 
The Josephson current at $\varphi=\pi-0^+$ flows through such highly degenerate resonant states.
Our analysis provides a mathematical background for understanding 
the fractional Josephson effect in a dirty SNS junction which was
numerically shown in the previous papers~\cite{ya06,ya062}.

\section{Phenomenological theory}
The fractional Josephson effect in one-dimensional SIS
can be phenomenologically explained in terms of the effective hopping 
Hamiltonian between 
the two Majorana bound states.
At the edge of isolated semi-infinite $p$-wave superconductor, 
the electron operators at the edges are described by
\begin{eqnarray}
\Psi_L =  \gamma_L, \\
\Psi_R = \rmi \rme^{\rmi\varphi/2} \gamma_R,
\end{eqnarray}
where $\gamma_L$ ($\gamma_R$) is the operator of a Majorana fermion at the edge 
of left (right) superconductor.
The tunneling Hamiltonian between the two edges becomes
\begin{eqnarray}
H_T &= - t \left[ \Psi_L^\dagger \Psi_R + \Psi_R^\dagger \Psi_L \right] \label{ph1}\\
&= - 2 \rmi t \cos(\varphi/2)\;  \gamma_L \; \gamma_R.
\end{eqnarray}
The expectation value of the tunneling Hamiltonian 
could be
\begin{eqnarray}
\left\langle H_T \right\rangle = - 2tC_0 \cos(\varphi/2),
\end{eqnarray}
where we assume that $\langle \rmi\gamma_L \; \gamma_R \rangle=C_0$ is a constant.
The Josephson current calculated as
\begin{eqnarray}
J = \frac{e}{\hbar} \partial_\varphi \left\langle H_T \right\rangle
 =\frac{e 2tC_0}{\hbar} \sin(\varphi/2),
\end{eqnarray}
describes the fractional current-phase relationship.
At $\varphi= \pi -0^+$, we obtain
\begin{eqnarray}
\left\langle H_T \right\rangle = -  tC_0\;  0^+,\\
J =\frac{e 2tC_0}{\hbar} \left( 1- \frac{(0^+)^2}{8} \right).
\end{eqnarray}
The Josephson current takes its maximum, whereas the 
amplitude of the tunneling Hamiltonian is proportional to $0^+$. 
The Josephson current at $\varphi=\pi-0^+$ flows as a result of the resonant transmission 
through the junction. Therefore the amplitude of the current is not proportional to the amplitude 
of the tunneling Hamiltonian. This argument is valid as far as $\varphi = \pi - 0^+$.
At $\varphi=\pi$, the tunneling Hamiltonian vanishes exactly, which leads to the absence 
of the Josephson current. In this way, the phenomenological argument using Eq.~(\ref{ph1}) 
is consistent with the microscopic theory of the fractional Josephson effect.

\section{Conclusion}

We have studied the stability of more than one Majorana Fermion appearing in a
two-dimensional superconductor/normal metal/superconductor (SNS) junction
in terms of chiral symmetry of Hamiltonian, where the two superconductors
are characterized by spin-triplet $p_x$-wave symmetry.
When the phase difference across the junction $\varphi$ is either 0 or $\pi$,
the Hamiltonian of the SNS junction preserves chiral symmetry.
At $\varphi=\pi$, the Majorana bound states (MBSs) in the normal metal
can retain their high degree of degeneracy at the zero energy even in the presence of the impurity scatterings
because all of the MBSs belong to the same chirality.
As a consequence, the resonant transmission of a Cooper pair via such highly degenerate MBSs
 carries the Josephson current at $\varphi=\pi-0^+$.
The physical picture obtained in this paper well explains
the persistence of the fractional current-phase relationship in a dirty SNS junction
which was numerically shown in previous papers.
We have also discussed a way of understanding the fractional current-phase
relationship derived from a phenomenological tunneling Hamiltonian of a Majorana Fermion.

\ack
The authors are grateful to Y.~Tanaka and S.~-I.~Suzuki for useful discussions.
This work was supported by gTopological Materials Scienceh (No. 15H05852)
and KAKENHI (Nos. 26287069 and 15H03525) from the Ministry of Education, Culture, Sports,
Science and Technology (MEXT) of Japan, and by the Ministry of Education and Science of the
Russian Federation (Grant No. 14Y.26.31.0007).

\appendix
\section{Wave function of zero energy states}
\label{sec:zs}
We derive the wave functions of the zero-energy states appearing
in the two semi-infinite $p_x$-wave superconductors illustrated in Fig.~\ref{fig:junct}.
In the $y$ direction, the width is denoted by $W$ and the hard-wall boundary condition is applied.
The Bogoliubov-de Gennes equation is given by
\begin{eqnarray}
H_{\alpha}\varphi_{\alpha,E} = E \varphi_{\alpha,E}, \label{eq:bdgiso}\\
H_{\alpha} =
 \left[
\begin{array}{cc}
\xi (\bi{r})
&  \frac{ \Delta \rme^{\rmi\varphi_{\alpha}} }{k_\mathrm{F}} \partial_x \\
-\frac{ \Delta \rme^{-\rmi\varphi_{\alpha}} }{k_\mathrm{F}} \partial_x
& - \xi (\bi{r}) \\
\end{array}
\right] ,
\end{eqnarray}
where the index $\alpha =L$,$R$ labels the left superconductor ($x\leq -L_x$)
and the right superconductor  ($x\geq L_x$), respectively.
The superconducting phase is given as
\begin{eqnarray}
\varphi_L = 0, \\
\varphi_R = \varphi.
\end{eqnarray}
The Hamiltonian $H_{\alpha}$  preserves chiral symmetry as
\begin{eqnarray} 
(\rme^{\rmi\varphi_{\alpha} T_3} \Gamma)\;  H_{\alpha} \; (\rme^{\rmi\varphi_{\alpha} T_3} \Gamma)^{-1}= - H_{\alpha},\\
\Gamma =\left[
\begin{array}{cc}
0 & 1 \\
1 & 0 \\
\end{array} \right] = T_1, \quad 
T_3 =\left[
\begin{array}{cc}
1 & 0 \\
0 & -1 \\
\end{array} \right].
\end{eqnarray}
By solving Eq.~(\ref{eq:bdgiso}),
we obtain the wave function belonging to an energy $E$ as

\begin{eqnarray}
\psi_{\alpha,E} = \phi_n (x) Y_n(y)
\end{eqnarray}
\begin{eqnarray}
\phi_n (x) &= 
c_1 \left[ \begin{array}{cc}
E + \Omega_+ \\ -\rmi \Delta \rme^{-\rmi \varphi_{\alpha}} (k_+/k_\mathrm{F}) \\
\end{array}
\right] \rme^{\rmi k_+ x} \nonumber\\
&+ c_2 \left[ \begin{array}{cc}
\rmi \Delta \rme^{\rmi \varphi_{\alpha}} (k_-/k_\mathrm{F}) \\E + \Omega_-  \\
\end{array}
\right] \rme^{\rmi k_- x}  \nonumber\\
&+ c_3 \left[ \begin{array}{cc}
-\rmi \Delta \rme^{\rmi \varphi_{\alpha}} (k_-/k_\mathrm{F}) \\E + \Omega_-  \\
\end{array}
\right] \rme^{-\rmi k_- x}  \nonumber\\
&+ c_4 \left[ \begin{array}{cc}
E + \Omega_+ \\ \rmi \Delta \rme^{-\rmi \varphi_{\alpha}} (k_+/k_\mathrm{F}) \\
\end{array}
\right] \rme^{-\rmi k_+ x},
\end{eqnarray}
\begin{eqnarray}
Y_n(y) = \sqrt{\frac{2}{W}} \sin \left( \frac{n \pi}{W} y \right),\\
\Omega_{\pm} = \sqrt{E^{2}-\frac{\Delta^2}{\mu} \left( \mu_n-\frac{\Delta^2}{4\mu} \right)} \mp \frac{\Delta^2}{2\mu}, \\ 
k_{\pm} = k_n \sqrt{1 \pm \frac{\Omega_{\pm}}{\mu_n}},\qquad 
k_n = \frac{\sqrt{2 m \mu_n} } {\hbar}, \\
\mu_n= \mu-\frac{\hbar^2}{2m} \left( \frac{n \pi}{W} \right)^2,
\end{eqnarray}
where $c_i$ ($i=1$-$4$) are numerical coefficients.
At $E=0$,  the wave function is deformed as
\begin{eqnarray}
\psi_{\alpha,0} (x,y) = \left[\phi_{\alpha,n,1} (x) + \phi_{\alpha,n,2} (x) \right] Y_n(y), \label{eq:phi0}\\
\phi_{\alpha,n,1} (x) = \left[
\begin{array}{cc} \rme^{\rmi \frac{\varphi_{\alpha}}{2}} \\ -\rme^{-\rmi \frac{\varphi_{\alpha}}{2}}
\end{array}
\right] \left[
c_1^{\prime} \rme^{\rmi q_n x} + c_3^{\prime} \rme^{-\rmi q_n x }\right] \rme^{-x/\xi}, \label{eq:phi0m}\\
\phi_{\alpha,n,2} (x) = \left[
\begin{array}{cc} \rme^{\rmi \frac{\varphi_{\alpha}}{2}} \\ \rme^{-\rmi \frac{\varphi_{\alpha}}{2}}
\end{array}
\right] \left[
c_2^{\prime} \rme^{\rmi q_n x} + c_4^{\prime} \rme^{-\rmi q_n x }\right] \rme^{x/\xi},  \label{eq:phi0p}\\
\quad q_n = \sqrt{ k_n^2-\xi^{-2}}, \qquad
\xi=\hbar^2 k_\mathrm{F}/m \Delta_0.
\end{eqnarray}
We note that the components of $\phi_{\alpha,n,1}$ and $\phi_{\alpha,n,2}$ are the eigenstates of the chiral symmetry operator as
\begin{eqnarray}
(\rme^{\rmi\varphi_{\alpha} T_3} \Gamma) \phi_{\alpha,n,1} (x) = - \phi_{\alpha,n,1} (x) , \\
(\rme^{\rmi\varphi_{\alpha} T_3} \Gamma) \phi_{\alpha,n,2} (x) = \phi_{\alpha,n,2} (x).
\end{eqnarray}
First, we calculate the wave function of the zero-energy states in the left superconductor. 
We apply the boundary condition in the $x$ direction as
\begin{eqnarray}
\psi_{L,0}  (-\infty,y) = \psi_{L,0}  (-L_x,y) = 0.
\end{eqnarray}
As a result, 
we obtain the two zero-energy states for each propagating channel as
\begin{eqnarray}
\psi_{L} =   \frac{1}{\sqrt{N_n}}\left[
\begin{array}{cc} 1 \\ 1
\end{array}
\right]  {\rm sin}[ q_n (x + L_x)] \rme^{ x/\xi}Y_n(y),
\end{eqnarray}
where the normalization coefficient is denoted by $N_n$.
In is easy to show that the zero energy states of left superconductor
are simultaneously the  eigenstates of chiral symmetry operator $\Gamma$
with the eigenvalue $\gamma=+1$.
Next, we consider the right superconductor. 
By applying the boundary condition in the $x$ direction as
\begin{eqnarray}
\psi_{R,0}  (L_x,y) = \psi_{R,0}  (\infty,y) = 0,
\end{eqnarray}
we find the wave function for the zero-energy states as
\begin{eqnarray}
\psi_{R} &=   \frac{1}{\sqrt{N_n}}\left[
\begin{array}{cc} \rme^{\rmi \frac{\varphi}{2}} \\ - \rme^{-\rmi \frac{\varphi}{2}}
\end{array}
\right]  {\rm sin}[ q_n (x - L_x)] \rme^{- x/\xi}Y_n(y).
\end{eqnarray}
The zero-energy states of the right superconductor $\psi_{R}$ hold $\gamma=-1$ for
the chiral symmetry operator $\rme^{\rmi\varphi T_3} \Gamma$.



\Bibliography{37}
\bibitem{mj37} Majorana E 1937 {\it Nuovo Cimento} {\bf 14} 171
\bibitem{rg00} Read N and Green D 2000 {\it Phys. Rev. B} {\bf 61} 10267
\bibitem{yk01} Kitaev A Y 2001 {\it Phys. Usp.} {\bf 44} 131
\bibitem{fk08} Fu L and Kane C L 2008 {\it Phys. Rev. Lett.} {\bf 100} 096407
\bibitem{ms09} Sato M, Takahashi Y and Fujimoto S 2009 {\it Phys. Rev. Lett.} {\bf 103} 020401
\bibitem{js10} Sau J D, Lutchyn R M, Tewari S, and DasSarma S 2010 {\it Phys. Rev. Lett.} {\bf 104} 040502
\bibitem{ja10} Alicea J 2010 {\it Phys. Rev. B} {\bf 81} 125318
\bibitem{rl10} Lutchyn R M, Sau J D and DasSarma S 2010 {\it Phys. Rev. Lett.} {\bf 105} 077001
\bibitem{yo10} Oreg Y, Refael G, and von Oppen F 2010 {\it Phys. Rev. Lett.} {\bf 105}, 177002
\bibitem{jy13} You J, Oh C H and Vedral V 2013 {\it Phys. Rev. B} {\bf 87} 054501
\bibitem{tc11} Choy T -P, Edge J M, Akhmerov A R and Beenakker C J W 2011 {\it Phys. Rev. B} {\bf 84} 195442
\bibitem{np13} Nadj-Perge S, Drozdov I K, Bernevig B A and Yazdani A 2013 {\it Phys. Rev. B} {\bf 88} 020407.
\bibitem{di01} Ivanov D A 2001 {\it Phys. Rev. Lett.} {\bf 86} 268
\bibitem{js11} Sau J D, Clarke D J and Tewari S 2011 {\it Phys. Rev B} {\bf 84} 094505
\bibitem{mourik} Mourik V, Zuo K, Frolov S M, Plissard S R, Bakkers E P A M and Kouwenhoven L P 2012 {\it Science} \textbf{336} 1003
\bibitem{deng} Deng M T, Yu C L, Huang G Y, Larsson M, Caroff P and Xu H Q 2012 {\it Nano Lett.} \textbf{12} 6414
\bibitem{yazdani} Nadj-Perge S, Drozdov I K, Li J, Chen  H ,Jeon S, Seo J, MacDonald A H, 
Bernevig B A and Yazdani A 2014 {\it Science} {\bf 346} 6209
\bibitem{kwon} Kwon H -J, Sengupta K and Yakovenko V M 2004 {\it Eur.\ Phys.\ J.~B} \textbf{37} 349$-$361
\bibitem{ishii} Ishii C 1970 {\it Prog. Theor. Phys.} \textbf{44} 1525
\bibitem{ishii2} Ishii C 1972 {\it Prog. Theor. Phys.} \textbf{47} 1646
\bibitem{ko2} Kulik I O and Omel'yanchuk A N, 1977 {\it Sov. J. Low. Temp. Phys.} \textbf{3} 459
[1977 {\it Fiz. Nizk. Temp.} \textbf{3} 945 ] 
\bibitem{likharev} Likharev K K 1979 {\it Rev.\ Mod.\ Phys.} \textbf{51} 101
\bibitem{aagolubov} Golubov A A, Kupriyanov M Y and ll'ichev E 2004 {\it Rev.\ Mod.\ Phys.} \textbf{76} 411
\bibitem{buchholts} Buchholtz L J and Zwicknagl G 1981 {\it Phys. Rev. B} \textbf{23} 5788
\bibitem{tanaka0} Tanaka Y and Kashiwaya S1995 {\it Phys. Rev. Lett.} \textbf{74} 3451
\bibitem{hu} Hu C R 1994 {\it Phys. Rev. Lett.} \textbf{72} 1526
\bibitem{ya04-1} Asano Y, Tanaka Y and Kashiwaya S 2004 {\it Phys. Rev. B} \textbf{69} 134501 
\bibitem{yt96-1} Tanaka Y and Kashiwaya S 1996 {\it Phys. Rev. B} \textbf{53} R11957
\bibitem{barash} Barash Y S, Burkhardt H and Rainer D 1996 {\it Phys. Rev. Lett.} \textbf{77} 4070
\bibitem{ya13} Asano Y and Tanaka Y 2013 {\it Phys. Rev B} {\bf 87} 104513
\bibitem{ya06} Asano Y, Tanaka Y, and Kashiwaya S 2006 {\it Phys.\ Rev.~Lett.} \textbf{96} 097007
\bibitem{ya062} Asano Y, Tanaka Y, Yokoyama T and Kashiwaya S 2006 {\it Phys.\ Rev.~B} \textbf{74} 064507
\bibitem{tewari} Tewari S and Sau J D 2012 {\it Phys. Rev. Lett.} \textbf{109} 150408
\bibitem{niu} Niu Y, Chung S -B, Hsu C -H, Mandal I, Raghu S and Chakravarty S 2012 {\it Phys. Rev. B} \textbf{85} 035110
\bibitem{diez} Diez M, Dahlhaus J P, Wimmer M and Beenakker C W J 2012 {\it Phys. Rev. B} \textbf{86} 094501
\bibitem{si15} Ikegaya S, Asano Y and Tanaka Y, 2015 {\it Phys.\ Rev.~B} \textbf{91} 174511
\bibitem{ms11} Sato M, Tanaka Y, Yada K and Yokoyama T 2011 {\it Phys. Rev. B} {\bf 83} 224511
\endbib
\end{document}